\documentclass[twocolumn,prb,preprintnumbers,amsmath,amssymb,floats,citeautoscript,nobalancelastpage,showpacs]{revtex4}
\usepackage{here}
\usepackage{graphicx}% Include figure files
\usepackage{dcolumn}% Align table columns on decimal point
\usepackage{bm}
\usepackage{color}
\usepackage{times}
\usepackage{multirow}

\def\vector#1{\mbox{\boldmath $#1$}}

\begin{document}

\title{Spin Dynamics in a Stripe-ordered Buckled Honeycomb Lattice Antiferromagnet Ba$_{2}$NiTeO$_{6}$}

\author{Shinichiro Asai$^1$}
\author{Minoru Soda$^1$}
\author{Kazuhiro Kasatani$^2$}
\author{Toshio Ono$^2$}
\author{V. Ovidiu Garlea$^3$}
\author{Barry Winn$^3$}
\author{Takatsugu Masuda$^1$}

\affiliation{$^1$Institute for Solid State Physics, The University of Tokyo, Kashiwanoha, Kashiwa, Chiba 277-8581, Japan \\
$^2$Department of Physical Science, Osaka Prefecture University, Sakai, Osaka 599-8531, Japan \\
$^3$Quantum Condensed Matter Division, Oak Ridge National Laboratory, Oak Ridge, Tennessee 37831, USA}

\begin{abstract}
We carried out inelastic neutron scattering experiments on 
a buckled honeycomb lattice antiferromagnet Ba$_{2}$NiTeO$_{6}$ exhibiting a stripe structure at a low temperature. 
Magnetic excitations are observed in the energy range of $\hbar \omega \lesssim 10$ meV 
having an anisotropy gap of 2 meV at 2 K.
We perform spin-wave calculations to identify the spin model.
The obtained microscopic parameters are consistent with the location of the stripe structure in the 
classical phase diagram. 
Furthermore, the Weiss temperature independently estimated from a bulk magnetic susceptibility is 
consistent with the microscopic parameters. 
The results reveal that a competition between the NN and NNN interactions that together with a relatively large single ion magnetic anisotropy stabilize the stripe magnetic structure.
\end{abstract}


\maketitle

\section{Introduction}
The honeycomb lattice antiferromagnets have attracted great interests in the geometrically frustrated magnets. 
Even though the simple $\rm{N\acute{e}el}$ order is the classical ground state for the plain system, 
the introduction of further neighbor interactions induces magnetic frustration and leads to various ordered states 
including spiral and stripe structures~\cite{PhysicaB971,EPJB20241}.
In case of the quantum spin case, a novel type of disordered state called 
plaquette valence-bond crystal is predicted~\cite{JPCM23226006,PRB84024406}.
Furthermore, the quantum spin liquid is suggested in the exactly solvable Kitaev model~\cite{Kitaev20062}, 
which is realized in the anisotropic Ising model on the plain honeycomb lattice~\cite{PRL102017205}. 
From the viewpoint of experiments, several magnetic states have been identified in regular honeycomb lattice antiferromagnets.
The $\rm{N\acute{e}el}$ order appears for the quasi-two-dimensional antiferromagnets BaNi$_{2}$V$_{2}$O$_{8}$ and BaNi$_{2}$P$_{2}$O$_{8}$~\cite{PRB65144443,JMMM151021}. 
Spiral magnetic order is stabilized in the isostructural compound BaCo$_{2}$As$_{2}$O$_{8}$~\cite{PhysicaBC86660}.
The spin-glass like disorder emerges in zero magnetic field in bilayer honeycomb lattice antiferromagnet Bi$_{3}$Mn$_{4}$O$_{12}$(NO$_{3}$)~\cite{PRL105187201}. 
The zigzag magnetic orders were observed in the single-layer honeycomb lattice antiferromagnet Na$_{2}$Co$_{2}$TeO$_{6}$~\cite{PRB94214416} and the Kitaev model compound $\alpha $-RuCl$_{3}$~\cite{PRB91144420}.
On the other hand, the stripe order, which was theoretically predicted, 
had not been found in a real compound until we reported on our previous study~\cite{PRB93024412} in Ba$_{2}$NiTeO$_{6}$.

\begin{figure}[h!]
\begin{center}
\includegraphics[width=70.0mm,clip]{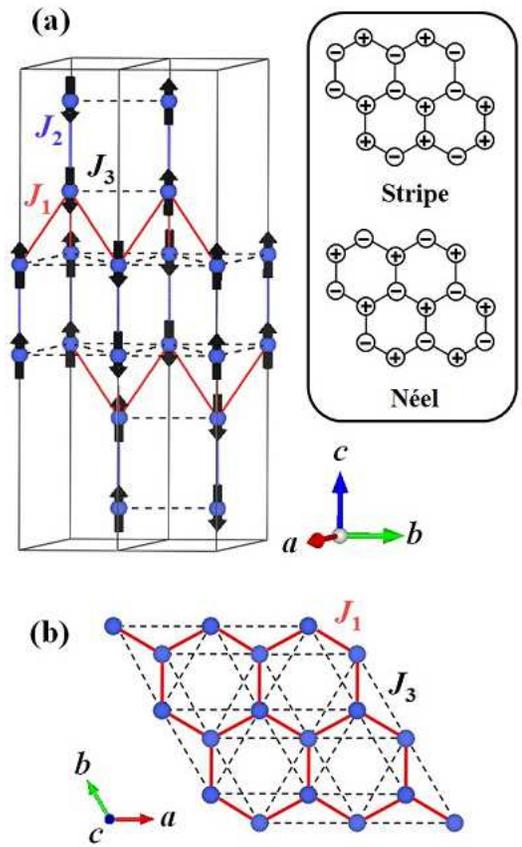}\\
\caption{
(Color online) (a) Magnetic structure of Ba$_{2}$NiTeO$_{6}$~\cite{PRB93024412}.
The blue spheres represent the Ni$^{2+}$ ions. 
The red solid, blue solid, and black dotted lines represent the pathways of \textit{J}$_{1}$, \textit{J}$_{2}$, and \textit{J}$_{3}$, respectively.
VESTA software~\cite{refforvesta} is used for drawing magnetic structure.
Inset shows spin arrangement of stripe and $\rm{N\acute{e}el}$ structure.
(b) Arrangement of Ni$^{2+}$ ions in buckled honeycomb lattice.
}
\label{fig1}
\end{center}
\end{figure}

Ba$_{2}$NiTeO$_{6}$ is a rare experimental realization 
of the buckled honeycomb lattice antiferromagnet~\cite{Ba2NiTeO6struc}.
The magnetic Ni$^{2+}$ ions and the pathways of their interactions \textit{J}$_{1}$, \textit{J}$_{2}$, and 
\textit{J}$_{3}$ are shown in Fig.\ 1(a).
The two neighboring triangular lattices coupled by the first-neighbor (NN) interaction \textit{J}$_{1}$ form a buckled honeycomb lattice as shown in Fig.\ 1(b)~\cite{PRB93024412}.
The third-neighbor interaction \textit{J}$_{3}$ corresponds to the next-nearest neighbor (NNN) interaction in the 
honeycomb lattice.
The buckled honeycomb lattices are magnetically coupled by the second-neighbor interaction \textit{J}$_{2}$.
Since the \textit{J}$_{1}$ and \textit{J}$_{3}$ have the similar Ni$^{2+}$-O$^{2-}$-O$^{2-}$-Ni$^{2+}$ paths 
owing to the buckled geometry of the honeycomb lattice,
they are expected to be comparative and induce strong frustration.
The magnetic susceptibility and heat capacity measurements identified a magnetic transition at 8.6 K~\cite{PRB93024412}.
The strong magnetic frustration and/or low dimensionality is indicated from the large frustration parameter ${\theta}_{\rm W} /T_{\rm N} = 18.6$,
where ${\theta}_{\rm W}$ is Weiss temperature and $T_{\rm N}$ is the magnetic transition temperature. 
We note that Ba$_{2}$CoTeO$_{6}$ also includes buckled honeycomb layers~\cite{DT395490}, 
and exhibited interesting magnetic phases 
particularly in the magnetic field~\cite{PRB93094420}. 
The material is, however, composed of two subsystems; a buckled honeycomb lattice and a triangular lattice. 

Recently, we investigated the magnetic structure of Ba$_{2}$NiTeO$_{6}$,
and we found that the collinear stripe structure 
with the propagation vector $\vector{k_{\bf mag}}$ of (0, 0.5, 1) 
is realized as shown in Fig.\ 1~\cite{PRB93024412}.
We classically calculated the phase diagram of $D/J_1$ vs $J_3/J_1$ for the buckled 
honeycomb lattice antiferromagnet. 
Here $D$ is the easy-axis type single-ion anisotropy. 
We demonstrated the existence of the stripe structure that is stabilized by a subtle balance between \textit{J}$_{1}$, \textit{J}$_{3}$, and $D$. 

In this paper, we investigate the spin dynamics of the titled compound 
by using the inelastic neutron scattering technique to identify the spin Hamiltonian. 
We observed a magnetic excitation having an energy gap at a low temperature. 
The neutron spectrum is reasonably reproduced by the calculation using linear spin-wave theory. 
The obtained set of parameters of exchange interaction and single-ion anisotropy are consistent with 
the location of the stripe structure in the phase diagram of the buckled honeycomb antiferromagnet. 

\section{Experimental Details}
The polycrystalline sample was synthesized by the solid state reaction method~\cite{Onounpublished}.
Inelastic neutron scattering measurements were performed at the hybrid spectrometer 
HYSPEC at the Spallation Neutron Source at Oak Ridge National Laboratory~\cite{HYSPECref}.
The incident neutron energies $E_i$ of 7.5, 15, and 35 meV were independently used.
For each of these energies the Fermi chopper frequency was set to be 300 Hz. 
The full width of the (015) nuclear peak at half maximum along the energy 
transfer ($\hbar \omega $) direction is evaluated to be 0.25(1), 0.66(3), and 1.95(5) meV for 
$E_i$ = 7.5, 15, and 35 meV, respectively.   
The low temperatures were achieved by the ORANGE cryostat.

\section{Results and Analysis}

\begin{figure*}
\begin{center}
\includegraphics[width=140.0mm,clip]{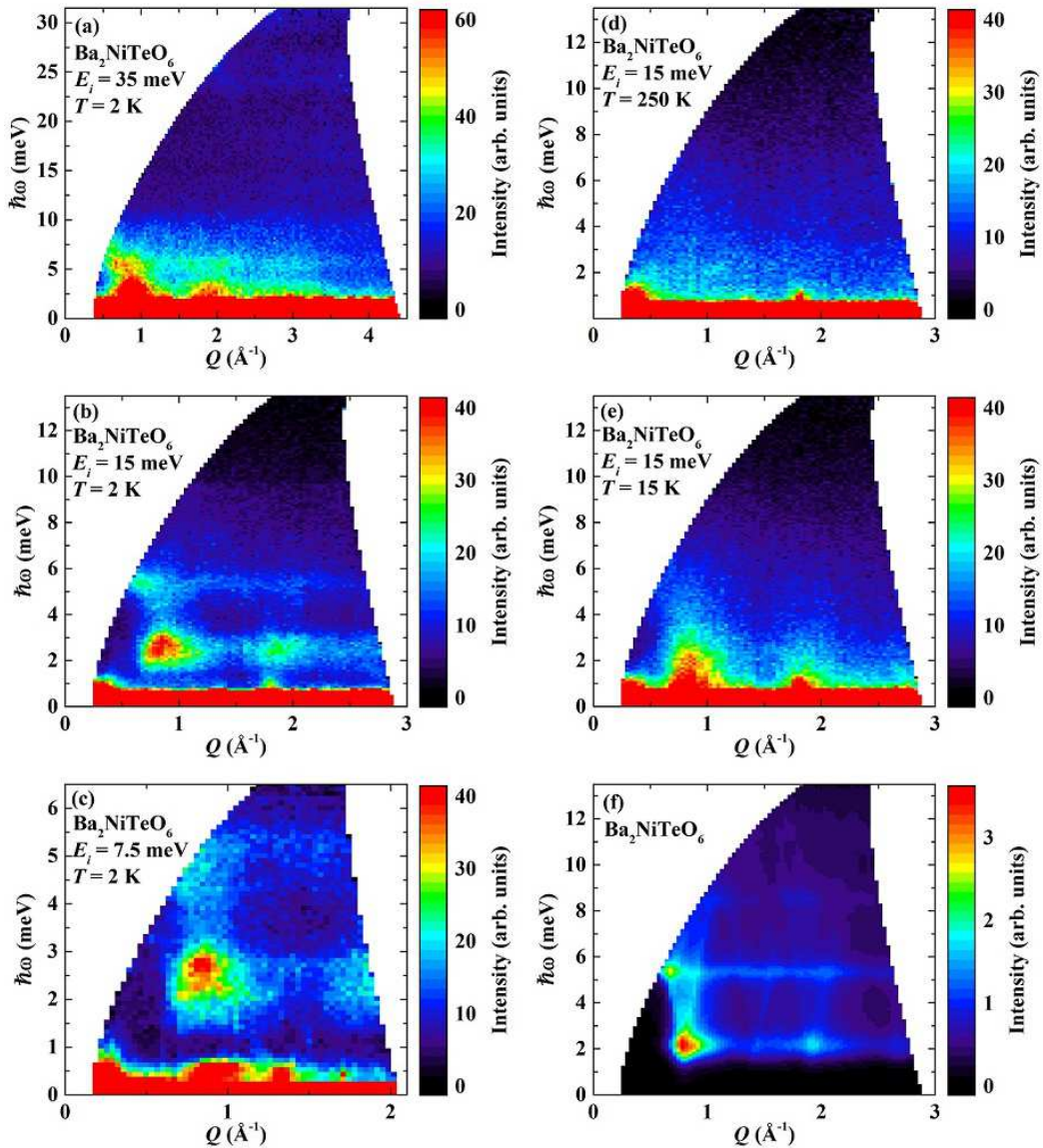}\\
\caption{
(Color online) INS spectra at 2 K for $E_{i}$ = (a) 35, (b) 15, and (c) 7.5 meV. INS spectra for $E_{i}$ = 15 meV at (d) 250 and (e) 15 K. 
(f) Calculated spin-wave spectra with $J_{1}$ = 0.8, $J_{2}$ = -0.01, $J_{3}$ = 1.6, $D$ = 1.1 meV.}
\label{fig2}
\end{center}
\end{figure*}

Figure 2(a), 2(b), and 2(c) show the inelastic neutron scattering (INS) spectra at 2 K for $E_{i}$ = 35, 15, and 7.5 meV, respectively. 
In Fig.~\ref{fig2}(a) excitations having strong intensities are observed at 
$\hbar \omega \lesssim 10$ meV. 
They decrease with the increase of $Q$, meaning that the dominant component is magnetic scattering. 
At $\hbar \omega \sim $ 13, 18, and 25 meV smeared and weak intensities are observed; 
all of them slightly increase with $Q$, meaning that they are not magnetic excitations. 
The energy band of the magnetic excitation is, thus, 10 meV. 
In Figs.~\ref{fig2}(b) and \ref{fig2}(c) the structure of the magnetic excitation 
is clearly observed. 
It exhibits an energy gap of 2 meV.
There are flat features at $\hbar \omega $ = 2.5 and 5.0 meV in the spectrum.
In the former two broad maxima are observed at $Q$ = 0.8 and 1.8 \AA $^{-1}$. 
 
The INS spectrum for $E_{i}$ = 15 meV at 250 K is shown in Fig.\ 2(d).
The excitation observed at 2 K is suppressed, and no clear feature is observed. 
The smeared excitations are ascribed to paramagnetic spins. 
Figure 2(e) shows the spectrum at 15 K. 
Smeared dispersive excitations which indicates a short-range spin correlation are observed. 
It is consistent with a broad maxima observed in the magnetic susceptibility~\cite{PRB93024412}.

In order to evaluate the magnetic interactions and the anisotropy from the obtained magnetic excitation, we calculate neutron cross section using spin-wave approximation.
Here we consider the Heisenberg model with easy-axis anisotropy as the same as that used in the previous study~\cite{PRB93024412}, which is given by, 
\begin{equation}
H=\sum_{i,j}\textit{J}_{i,j} \vector{S}_{i}\cdot \vector{S}_{j}-D\sum_{i}S^{2}_{i,z},
\end{equation} 
where $\vector{S}_{i}$ and $S_{i,z}$ represent the vectors for the spin of Ni$^{2+}$ ion at the position of $\vector{r_{i}}$ and its component along the \textit{c} axis, respectively.
We take the sum in the first term of the Eq.\ (1) for all the pairing of spins corresponding to \textit{J}$_{1}$, \textit{J}$_{2}$, and \textit{J}$_{3}$ in the unit cells.
We assume the stripe structure determined by our previous study~\cite{PRB93024412} as the ground state.
The cross section for this model is calculated by the method described in Ref.\ 20.
We take a powder average in order to compare the calculated spectrum with the experimentally obtained one.
The calculated spectrum is convoluted by the Gaussian function.
The resolution along $Q$ direction is experimentally obtained from the width of the nuclear (015) peak.
The resolution along $\hbar \omega $ direction $d(\hbar \omega )$ is roughly approximated to be $d(\hbar \omega ) = 0.4789-0.032\hbar \omega $ meV as a function of $\hbar \omega $, which is evaluated from the linear fitting of the instrumental resolution.
In order to reproduce the width of the broad peaks in the spectrum, we further convolute the spectrum along the $\hbar \omega $ direction by the Lorentzian function with the width of the Lorentzian function $dL$ of 0.15 meV,
which indicates a slight decrease of life time of the excitation. 
We found that $J_{2}$ is needed to be much smaller than $J_{1}$, $J_{3}$, and $D$ in 
order to reproduce the flat components in experimentally obtained spectra. 
$J_{2}$ is negligibly small so that it cannot be evaluated qualitatively.
We put negative small value (-0.01 meV) in $J_{2}$ to ensure the stripe structure in the previous study~\cite{PRB93024412}.
The parameters are thus $J_{1}$, $J_{3}$, $D$.
We finally determined the parameters to be $J_{1}$ = 0.8 (1), $J_{3}$ = 1.6(1), and $D$ = 1.1(1) meV.
The calculated spin-wave spectrum with the parameters $J_{1}$ = 0.8, $J_{3}$ = 1.6, $D$ = 1.1, and $dL$ = 0.15 meV is shown in Fig.\ 2(f).
It reproduces the experimentally obtained spectrum shown in Fig.\ 2(b).

\begin{figure}
\begin{center}
\includegraphics[width=60.0mm,clip]{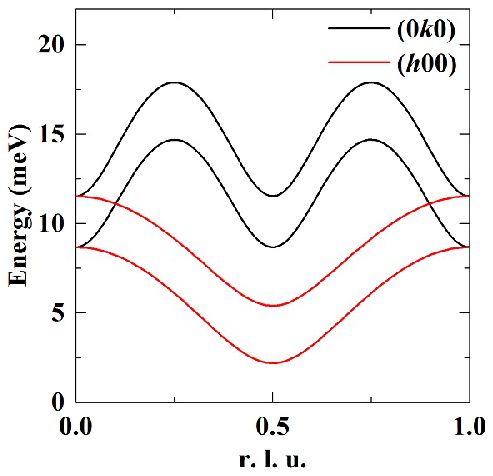}\\
\caption{
In-plane dispersion for the calculated spin-wave spectra with $J_{1}$ = 0.8, $J_{2}$ = -0.01, $J_{3}$ = 1.6, $D$ = 1.1 meV.
Red and black solid lines show the dispersions along the $h$ and $k$ directions in the reciprocal lattice space.
}
\end{center}
\end{figure}

\begin{figure}
\begin{center}
\includegraphics[width=80.0mm,clip]{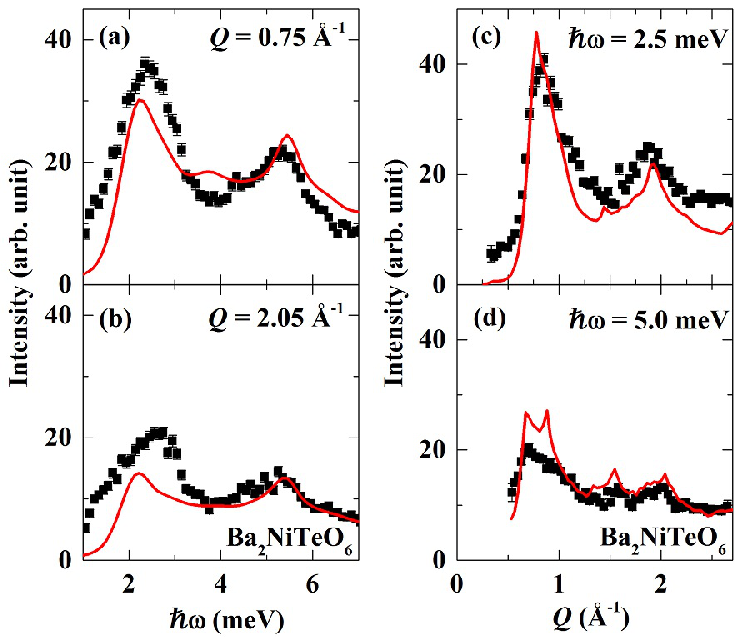}\\
\caption{
(Color online)
(a) One-dimensional cut along energy transfer at $Q$ = 0.75 \AA $^{-1}$ obtained by integrating intensity with respect to $Q$ from 0.7 to 0.8 \AA $^{-1}$.
(b) One-dimensional cut along energy transfer at $Q$ = 2.05 \AA $^{-1}$ obtained by integrating intensity with respect to $Q$ from 2.0 to 2.1 \AA $^{-1}$.
(c) One-dimensional cut along $Q$ at $\hbar \omega $ = 2.5 meV obtained by integrating intensity with respect to $\hbar \omega $ from 2.3 to 2.7 meV.
(d) One-dimensional cut along $Q$ at $\hbar \omega $ = 5.0 meV obtained by integrating intensity with respect to $\hbar \omega $ from 4.8 to 5.2 meV.
Square symbols represent the experimental data while the curves depict the calculated values.
Incoherent elastic component is subtracted from the the experimental data.
The detail is shown in the main text.
}
\end{center}
\end{figure}

Here we discuss the detail of the calculated spectrum.
We show the in-plane dispersions in Fig.\ 3.
The difference of the period for the dispersions between the $h$ and $k$ directions corresponds to the modulation of the stripe order.
The bottom of the dispersion along the $h$ direction is located at 2.2 and 5.4 meV.

Let us compare the one-dimensional cuts of the spectra shown in Fig.\ 2(b) and 2(f) in order to compare them more in detail.
In order to exclude the incoherent elastic component from the experimental data,
we evaluate it by fitting the peak at $\hbar \omega $ = 0 meV in the one-dimensional cut along energy transfer obtained by integrating intensity with respect to $Q$ from 0.7 to 0.8 by using Gaussian function,
and subtracted it from the one-dimensional cuts shown in Fig.\ 4(a), 4(b), 4(c), and 4(d).  
Figure 4(a) and 4(b) show the one-dimensional cuts along energy transfer at $Q$ = 0.75 and 2.05 \AA $^{-1}$ obtained by integrating intensity with respect to $Q$ from 0.7 to 0.8 and from 2.0 to 2.1 \AA $^{-1}$, respectively.
One-dimensional cuts along $Q$ at $\hbar \omega $ = 2.5 and 5.0 meV are shown in Fig.\ 4(c) and 4(d), which are obtained by integrating intensity with respect to $\hbar \omega $ from 2.3 to 2.7 and from 4.8 to 5.2 meV, respectively.
We clearly see that the calculation roughly reproduces the broad peaks along $Q$ direction observed at $\hbar \omega $ = 2.5 meV and those along $\hbar \omega $ direction at $Q$ = 0.75 \AA $^{-1}$.
More precise estimate of the magnetic interactions and anisotropy is needed for better reproduction, which can be achieved by the single-crystal neutron scattering study.
%We thus conclude that the observed magnetic excitation is well explained by the spin-wave calculation.

\section{Discussion}

\begin{figure}
\begin{center}
\includegraphics[width=85.0mm,clip]{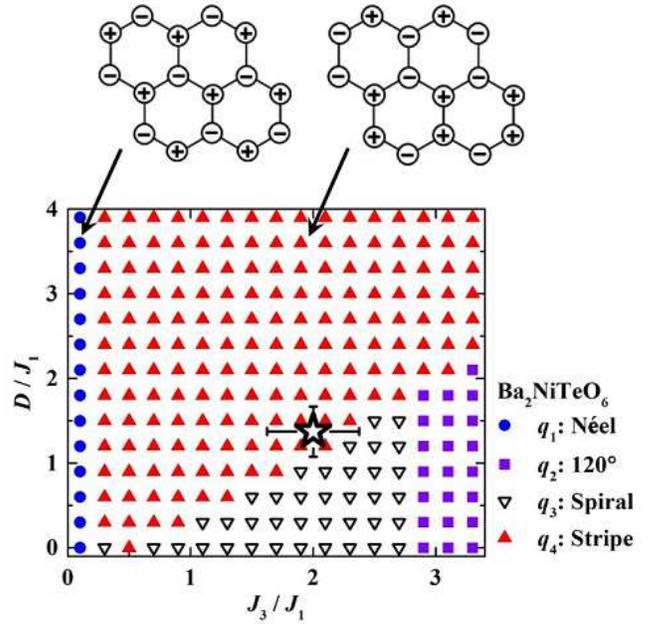}\\
\caption{
Magnetic phase diagram for classical ground state.
The $\vector{q}_{1}$ and $\vector{q}_{4}$ phases are sketched above the diagram.
The $\vector{q}_{4}$ phase corresponds to the experimentally determined magnetic structure.
The details of the $\vector{q}_{1}$, $\vector{q}_{2}$, $\vector{q}_{3}$, and $\vector{q}_{4}$ phases are described in Ref.\ 13.
}
\end{center}
\end{figure}

We first discuss the evaluated parameters of magnetic interactions and anisotropy in the phase diagram in the previous study~\cite{PRB93024412}.
Figure~5 shows the classical phase diagram in the range of $0 < J_{3}/J_{1} < 3.2$ and $ 0 < D/J_{1} < 4$.
N{\'e}el, 120 degree, and spiral structures are realized in the $\vector{q}_{1}$, $\vector{q}_{2}$, $\vector{q}_{3}$ phases, respectively.
In the $\vector{q}_{4}$ phase, the stripe structure is stabilized.
We note that the $\vector{q}_{4}$ phase is only stable at $J_{3}/J_{1} = 0.5$ in the case of $D = 0$~\cite{EPJB20241}. 
The $J_3$ induces geometrical frustration and supports spiral structure. 
The easy-axis anisotropy, on the other hand, suppresses the spiral structure and stabilizes 
the stripe structure. 
This means that the stripe structure appears as the result of competition between geometrical 
frustration in 
the buckled honeycomb lattice and the easy-axis anisotropy. 
The open star symbol in the phase diagram represents the parameters obtained in the present experiment.
The symbol is located in the region of stripe structure, 
meaning that the obtained parameters from the INS experiment 
are consistent with the classical phase diagram of the ground state.

Next, let us discuss the magnetic behavior above $T_{\rm N}$ of this compound.
The small $J_{2}$ suggests that the buckled honeycomb lattices in Ba$_{2}$NiTeO$_{6}$ are magnetically isolated.
The strong magnetic frustration is expected from the comparative magnetic interactions $J_{1}$ and $J_{3}$.
These obtained results are consistent with the broad maximum of the magnetic susceptibility and heat capacity indicating the short-range magnetic correlation of Ni$^{2+}$ ions~\cite{PRB93024412}.
The magnetic susceptibility of the powder sample follows the Curie-Weiss law above 100 K.
We expects that the Weiss temperature depends on the magnetic interactions and anisotropy,
and it is a good indicator for examining whether the evaluated parameters 
are quantitatively consistent with the magnetic properties.
Then, let us derive the Weiss temperature analytically by the mean-field approximation.
We safely neglect the $J_{2}$ in the calculation because of its small value. 
We consider the Heisenberg Hamiltonian with the easy-axis anisotropy and applied magnetic field $\vector{H}$, which is given by,
\begin{equation}
\mathcal{H}=\sum_{i,j}\textit{J}_{i,j} \vector{S}_{i}\cdot \vector{S}_{j}-D\sum_{i}S^{2}_{i,z}+\sum_{i}g\mu _{\rm B}\vector{S}_{i}\cdot \vector{H}.
\end{equation} 
By using the mean-field approximation, the Hamiltonian is rewritten as
\begin{equation}
\mathcal{H}=-\sum_{i}DS^{2}_{i,z}+\sum_{i}g\mu _{\rm B}\vector{S}_{i}\cdot \vector{H}',
\end{equation} 
where $\vector{H}'$ is the sum of the molecular and applied magnetic field, which is given by
\begin{equation}
\vector{H}'=\vector{H}+\frac{\sum_{j}\textit{J}_{i,j}<\vector{S}_{j}>}{g\mu _{\rm B}}=\vector{H}+\frac{(3J_{1}+6J_{3})<\vector{S}_{i}>}{g\mu _{\rm B}}.
\end{equation}
For simplicity, we assume that the applied magnetic field is parallel to the crystallographic $c$ axis, which is the magnetic easy axis.
Then, the Hamiltonian is modified as
\begin{equation}
\mathcal{H}=-\sum_{i}DS^{2}_{i,z}+\sum_{i}g\mu _{\rm B}S_{i,z}H'.
\end{equation} 
In mean-field approximation the eigenenergy of the spin is independent of the position $\vector{r_{i}}$, 
and the energy is given by
\begin{equation}
\epsilon _{m}=-Dm^{2}+g\mu _{\rm B}mH',
\end{equation} 
where $m$ is the magnetic quantum number ($m$ = 1, 0, -1).
Then, we can evaluate $<S_{z}>$ by using the partition function expressed as,
\begin{equation}
<S_{z}>=\cfrac{\sum_{m}m \exp (\cfrac{-\epsilon _{m}}{k_{\rm B}T})}{\sum_{m} \exp (\cfrac{-\epsilon _{m}}{k_{\rm B}T})},
\end{equation} 
In the case of $k_{\rm B}T >> \mid\epsilon_{m}\mid $, we can approximate the exponential functions by the linear functions.
We substitute the equations (4) and (6) for the equation (7), and derive $<S_{z}>$ as
\begin{equation}
<S_{z}>=\cfrac{-2g\mu _{\rm B}H}{3k_{\rm B}T+2(3J_{1}+6J_{3}+D)}.
\end{equation} 
Then, we obtained the magnetic susceptibility following the Curie-Weiss law, which is given by
\begin{equation}
\chi =\cfrac{-g\mu _{\rm B}<S_{z}>}{H}=\cfrac{C}{T+\theta },
\end{equation}
where $C$ is the Curie constant.
We finally derive the Weiss temperature $\theta$ as
\begin{equation}
\theta =\cfrac{2(3J_{1}+6J_{3}+D)}{3k_{\rm B}}.
\end{equation} 
Substitute the parameters estimated from the INS experiment, 
$J_{1}$ = 0.8, $J_{3}$ = 1.6, $D$ = 1.1 meV, 
for the equation (10), and we obtained $\theta$ $\sim$ 101 K.
It is consistent with the experimentally obtained 
value of 128(2) K from the magnetic susceptibility for the single crystal sample
in the case that $\vector{H}$ is parallel to the $c$ axis~\cite{Onounpublished}. 
Thus, the consistency among the independent experiments, the INS and bulk properties measurements, is confirmed. 

Here we discuss the difference of the energy scale between the magnetic interactions $J_{1}$ and $J_{3}$.
The superexchange interactions via the  Ni$^{2+}$-O$^{2-}$-O$^{2-}$-Ni$^{2+}$ pathways for $J_{1}$ and $J_{3}$ are expected to be antiferromagnetic from the Goodenough-Kanamori rule~\cite{KanamoriGoodenoughRef}.
Additionally, the antiferromagnetic contribution from the Ni$^{2+}$-O$^{2-}$-Te$^{6+}$-O$^{2-}$-Ni$^{2+}$ pathways are also expected as discussed for Sr$_{2}$CuTeO$_{6}$~\cite{PRB93054426}. 
On the other hand, these Te$^{6+}$-mediated interactions favors $J_{1} > J_{3}$ because there are two pathways for $J_{1}$ in contrast to a single pathway for $J_{3}$, which is not consistent with the obtained results.
We expects that the result $J_{1} < J_{3}$ is due to the distortion of NiO$_{6}$ octahedra which change the Ni$^{2+}$-O$^{2-}$-O$^{2-}$-Ni$^{2+}$ bond angle.
Theoretical calculation of the magnetic interactions is needed for further investigation.

Finally we compare the magnetic property of honeycomb lattice for Ba$_{2}$NiTeO$_{6}$ with that for other honeycomb lattice antiferromagnets.
In Ba$_{2}$CoTeO$_{6}$ where the buckled honeycomb and triangular layers are alternately stacked,
the magnetic order in the honeycomb layer is similar to that of Ba$_{2}$NiTeO$_{6}$~\cite{DT395490}.
The ratio of the NNN interaction to the NN interaction in Ba$_{2}$CoTeO$_{6}$ is estimated to be about $1/4$ of that in Ba$_{2}$NiTeO$_{6}$~\cite{PRB93094420}.
This means that the magnetic interactions are much modified by the substitution of Ni$^{2+}$ ions for Co$^{2+}$ ions, and nevertheless, the stripe order is retained.
The result is consistent with our phase diagram in Fig.~5; the stripe order is robust to  $J_{3}/J_{1}$ 
under the existence of the the easy-axis type anisotropy, 
and particularly at $J_{3}/J_{1}$ = 0.5 solely the stripe order does exist regardless of the magnitude of $D$. 

In contrast with the buckled honycomb lattice magnets, 
the thrid-neighbor interaction is rather enhanced in the regular honeycomb lattice magnets 
formed by edge-shared octahedra of 
the ligands such as BaNi$_{2}$P$_{2}$O$_{8}$~\cite{JMMM151021} and BaCo$_{2}$As$_{2}$O$_{8}$~\cite{PhysicaBC86660}. 
The magnetic frustration induced by the competition between  the NN and third-neighbor interactions leads to 
various states, which had been studied in the accumulative studies~\cite{PRB65144443,JMMM151021,PhysicaBC86660,PRL105187201,PRB94214416,Regnault89}. 
Meanwhile the magnetic states of the buckled honeycomb lattice antiferromagnets has been less 
explored because of a small amount of the model compounds.
The investigation on other model compounds are needed for further study.

\section{Summary}
We carried out the INS experiment for the buckled honeycomb lattice 
antiferromagnet Ba$_{2}$NiTeO$_{6}$ in order to investigate 
the magnetic interaction and anisotropy quantitatively.
The magnetic excitation with the band energy of 10 meV and an 
energy gap of 2 meV is observed at 2 K.
We perform the spin-wave calculation and evaluate the magnetic interaction and anisotropy.
The spectrum at 2 K is well reproduced by the calculated one with 
the parameters $J_{1}$ = 0.8(1), $J_{3}$ = 1.6(1), $D$ = 1.1(1) meV, $J_{2}$ is negligibly small.
The evaluated parameters are located in the range of stripe phase in the phase diagram of the 
frustrated honeycomb antiferromagnet. 
They are consistent with the Weiss temperature independently estimated from the bulk magnetic property measurement. 
The consistency among the INS experiment, magnetic susceptibility measurement, and 
the calculation of the ground state reveals that Ba$_{2}$NiTeO$_{6}$ is an experimental realization 
of the two-dimensional honeycomb lattice antiferromanget and that the stripe structure is 
the result of the competition between the geometrical frustration of the lattice and 
the easy-axis anisotropy of Ni spins.

\begin{acknowledgements}
This research used resources at the Spallation Neutron Source, a DOE Office of Science User Facility operated by the Oak Ridge National Laboratory (IPTS-13701.1).
Travel expenses for the neutron scattering experiment were supported by US-Japan Cooperative Program for Neutron Scattering Experiments (proposal no.\ 2015-04).
\end{acknowledgements}

\end{document}